\documentclass[10pt,conference]{IEEEtran}

\usepackage{graphicx,color}
\usepackage{amsmath, amsfonts, amssymb, amsbsy}
\usepackage{algorithm}
\usepackage{enumerate}
\usepackage{algorithmic}
\usepackage[sort,compress]{cite}
\usepackage{epsfig}
\usepackage{epstopdf}

\newcommand{\beq}{\begin{equation}}
\newcommand{\eeq}{\end{equation}}

\newcommand{\lb}{\left(}
\newcommand{\rb}{\right)}
\newcommand{\lsb}{\left[}
\newcommand{\rsb}{\right]}

\newcommand{\dsum}{\displaystyle\sum}

\newcommand{\dmin}{\displaystyle\min}

\def\adots{\mathinner{\mskip0mu\raise0pt\vbox{\kern7pt\hbox{.}}\mskip3mu
          \raise4pt\hbox{.}\mskip3mu\raise8pt\hbox{.}\mskip0mu}}

\newcommand{\Nt}{{N_t}}
\newcommand{\Nr}{{N_r}}

\newcommand{\xh}{\widehat{x}}

\newcommand{\tr}{\mbox{tr}}

\newcommand{\bmh}{\bfh}

\usepackage{bm}

\newcommand{\bmy}{{\bm y}}
\newcommand{\bmv}{{\bm v}}

\newcommand{\bmF}{{\bm F}}
\newcommand{\bmG}{{\bm G}}
\newcommand{\bmH}{{\bm H}}
\newcommand{\bmQ}{{\bm Q}}
\newcommand{\bmR}{{\bm R}}
\newcommand{\bmP}{{\bm P}}
\newcommand{\bmf}{{\bm f}}
\renewcommand{\bmh}{{\bm h}}

\newcommand{\bmS}{{\bm S}}

\newcommand{\bmA}{{\bm A}}

\newcommand{\bmC}{{\bm C}}

\newcommand{\bmHh}{\widehat{\bmH}}
\newcommand{\bmI}{{\bm I}}
\newcommand{\bmD}{{\bm D}}

\IEEEoverridecommandlockouts

\newcommand{\CN}{{\cal CN}}
\newcommand{\E}{\mbox{E}}

\newcommand{\bit}{\begin{itemize}}
\newcommand{\eit}{\end{itemize}}

\newcommand{\hwt}{\widetilde{h}}

\newcommand{\htt}{\widetilde{\hwt}}

\newcommand{\bmg}{{\mathbf g}}
\renewcommand{\bmf}{{\mathbf f}}
\renewcommand{\bmh}{{\mathbf h}}
\renewcommand{\bmH}{{\mathbf H}}

\newcommand{\kbar}{\overline{k}}

\renewcommand{\Nt}{M}
\renewcommand{\Nr}{N}
\newcommand{\Np}{{L}}

\newcommand{\bmHb}{\overline{\bmH}}
\newcommand{\ibar}{\overline{i}}
\renewcommand{\bmA}{{\mathbf A}}

\newcommand{\bmHt}{\widetilde{\bmH}}

\newcommand{\bmB}{{\mathbf B}}

\renewcommand{\bmS}{{\mathbf S}}

\newcommand{\bmT}{{\mathbf T}}

\newcommand{\bmgb}{\overline{\bmg}}

\newcommand{\bmhb}{\overline{\bmh}}
\newcommand{\bmW}{{\mathbf W}}

\newcommand{\bmPsi}{{\bm \Psi}}

\newcommand{\bmlambda}{{\mathbf \lambda}}
\newcommand{\bmGb}{\overline{\bmG}}

\begin{document}

\title{Noncoherent Multi-User MIMO Communications using Covariance CSIT}

\author{
\IEEEauthorblockN{
Christo Thomas Kurisummoottil\IEEEauthorrefmark{3},
Wassim Tabikh\IEEEauthorrefmark{3}\IEEEauthorrefmark{5},
Dirk Slock\IEEEauthorrefmark{3},
Yi Yuan-Wu\IEEEauthorrefmark{5}
\vspace{1mm}}
\IEEEauthorblockN{
\IEEEauthorrefmark{3}EURECOM, Sophia-Antipolis, France, Email: \{Christo.Kurisummoottil,tabikh,slock\}@eurecom.fr\\
\IEEEauthorrefmark{5}Orange Labs, Issy-les-Moulineaux, France, Email: yi.yuan@orange.com
}}

\maketitle
\begin{abstract}
The Multi-User downlink, particularly in a Multi-Cell Massive MIMO setting, requires enormous amounts of instantaneous CSIT (Channel State Information at the Transmitter(s)), iCSIT. Here we focus on exploiting channel covariance CSIT (coCSIT) only. In particular multipath induced structured low rank covariances are considered that arise in Massive MIMO and mmWave settings, which we call pathwise CSIT (pwCSIT). The resulting non-Kronecker MIMO channel covariance structures lead to a split between the roles of transmitters and receivers in MIMO systems. For the beamforming optimization, we consider a minorization approach applied to the Massive MIMO limit of the Expected Weighted Sum Rate. Simulations indicate that the pwCSIT based designs may lead to limited spectral efficiency loss compared to iCSIT based designs, while trading fast fading CSIT for slow fading CSIT. We also point out that the pathwise approach may lead to distributed designs with only local pwCSIT, and analyze the sum rates for iCSIT and pwCSIT in the low and high SNR limits.
\end{abstract}

\section{Introduction}

In this paper, Tx may denote transmit/transmitter/{\linebreak}transmission and Rx may denote receive/receiver/reception.
Interference is the main limiting factor in wireless transmission.
Base stations (BSs) disposing of multiple antennas are able to serve multiple Mobile Terminals (MTs) simultaneously, which is called
Spatial Division Multiple Access (SDMA) or Multi-User (MU) MIMO. However, MU systems have precise requirements for Channel State Information
at the Tx (CSIT) which is more difficult to acquire than CSI at the Rx (CSIR).
Hence we focus here on the more challenging downlink (DL).

The recent development of Massive MIMO (MaMIMO) \cite{b1} opens new possibilities for increased system capacity while at the same time simplifying system design. We refer to \cite{b2} for a further discussion of the state of the art, in which
MIMO Interference Alignment (IA) requires global MIMO channel CSIT. Recent works focus on intercell exchange of only scalar quantities, at fast fading rate, as
also on two-stage approaches in which the intercell interference gets zero-forced (ZF). Also, massive MIMO in most works refers actually
to MU MISO.

Whereas the exploitation of covariance CSIT (coCSIT) may be beneficial, in a MaMIMO context it may quickly lead to high computational complexity and
estimation accuracy issues. Computational complexity may be reduced (and the benefit of coCSIT enhanced) in the case of low rank or related covariance  structure, but the use and tracking of subspaces may still be cumbersome. In the pathwise approach, these subspaces are very parsimoniously parameterized.
In a FDD setting, these parameters may even be estimated from the uplink (UL).  As opposed to the instantaneous channel CSIT (iCSIT), the pathwise CSIT (pwCSIT) is not affected by fast fading.

Massive MIMO makes the pathwise approach viable. Indeed, with enough antennas, pwCSIT by itself may allow zero forcing (ZF) \cite{b3}, which is of interest at high SNR. However, we are particularly concerned here
with maximum Weighted Sum Rate (WSR) designs accounting for finite SNR. ZF of all interfering links leads to significant reduction
of useful signal strength.
We briefly allude to the general case of Gaussian partial CSIT (paCSIT), in which
the combined availability of channel estimates (mean CSIT) and coCSIT can be exploited.
Such general paCSIT scenario can e.g. be particularized as in \cite{b4} to the case of
perfect iCSIT for intracell channels and pwCSIT for intercell channels.
This leads to 2-stage BF expressions, similar to hybrid beamforming.
The slow stage handles intercell interference, and is frequency-flat. 
It can be exploited also to separate the cells for channel estimation purposes.
In what follows we consider in more detail pwCSIT for all channels (both intercell and intracell).
Also, in this (as any) case of paCSIT, the WSR criterion needs to be modified.
We shall consider the Expected WSR (EWSR). Furthermore, we shall take advantage of a Massive MIMO setting to 
exploit a simple Massive EWSR limit that results from the law of large numbers.
This MaEWSR limit leads to a loss of all (narrowband) frequency-selectivity in the channel and also
leaves no utility for space-time coding, though this can be expected to bring some benefits.

\section{Streamwise IBC Signal Model}

We start with a per stream approach (which in the perfect CSI case would be equivalent to per user).
In an IBC formulation, one stream per user can be expected to be the usual scenario. In the development below, in the case of more than one stream per user,
treat each stream as an individual user.
So, consider again an IBC with $C$ cells with a total of $K$ users. 
We shall consider a system-wide numbering of the users. User $k$ is served by BS $b_k$.
The $\Nr_k\times 1$ received signal at user $k$ in cell $b_k$ is
\beq
\bmy_k\! =\! \underbrace{\bmH_{k,b_k}\, \bmg_k\, x_k}_{\mbox{signal}} + \!\!\!\underbrace{\sum_{\stackrel{i\neq k}{b_i=b_k}} \!\!\bmH_{k,b_k}\,\bmg_i\, x_i}_{\mbox{intracell interf.}} +
\!\!\underbrace{\sum_{j\neq b_k} \sum_{i: b_i=j} \! \bmH_{k,j}\,\bmg_i\, x_i}_{\mbox{intercell interf.}}\! + \!\bmv_k
\label{eqIBCM1}
\eeq
where $x_k$ is the intended (white, unit variance) scalar signal stream, 
$\bmH_{k,b_k}$ is the $\Nr_k\times\Nt_{b_k}$ channel from BS $b_k$ to user $k$. 
BS $b_k$ serves $K_{b_k}=\sum_{i:b_i=b_k}1$ users.
We considering a noise whitened signal representation so  that we get for the noise
$\bmv_k\sim\CN(0, I_{\Nr_k})$.
The $\Nt_{b_k}\times 1$ spatial Tx filter or beamformer (BF) is $\bmg_k$.
Treating interference as noise, user $k$ will apply a linear Rx filter $\bmf_k$ to maximize the signal power (diversity) while reducing
any residual interference that would not have been (sufficiently) suppressed by the BS Tx. The Rx filter output is $\xh_k = \bmf_k^H\bmy_k$.

\section{Max WSR with Perfect CSIT}
\label{KG}

Consider as a starting point for the optimization the weighted sum rate (WSR)
\beq
WSR = WSR(\bmg) = \sum_{k=1}^K u_k \;\ln\frac{1}{e_k}
\label{eqWSR1}
\eeq
where $\bmg$ represents the collection of BFs $\bmg_k$, the $u_k$ are rate weights, the $e_k=e_k(\bmg)$ are the Minimum Mean Squared Errors (MMSEs) for
estimating the $x_k$:
\beq\mbox{}\!\!
\begin{array}{l}
\dfrac{1}{e_k} \!\!= \!\!1 \!+\!\bmg_k^H\bmH_{k,b_k}^H \bmR_{\kbar}^{-1} \bmH_{k,b_k}\bmg_k\! =\! ( 1 \!-\!\bmg_k^H\bmH_{k,b_k}^H\bmR_{k}^{-1}\bmH_{k,b_k} \bmg_k)^{-1}\\[2mm]
\bmR_k =  \bmH_{k,b_k}\bmQ_k\,\bmH_{k,b_k}^H +   \bmR_{\kbar}\, ,\;  \bmQ_i = \bmg_i\bmg_i^H\, ,\\[1mm]
\bmR_{\kbar} =  \dsum_{i\neq k}\bmH_{k,b_i}\bmQ_i\,\bmH_{k,b_i}^H +   I_{\Nr_k}\, .
\end{array}
\label{eqWSR2}
\eeq
$\bmR_k$, $\bmR_{\kbar}$ are the total and interference plus noise Rx covariance matrices resp.
and $e_k$ is the MMSE obtained at the output $\xh_k =\bmf_k^H\bmy_k$ of the optimal (MMSE) linear Rx $\bmf_k$, 
\beq
\bmf_k = \bmR_k^{-1} \bmH_{k,b_k}\bmg_k = \bmR_k^{-1} \bmh_{k,k}\; .
\label{eqWSR3}
\eeq
The WSR cost function needs to be augmented with the power constraints
\beq
\sum_{k:b_k=j}\tr\{\bmQ_k\} \leq P_j \, .
\label{eqWSR33}
\eeq

\subsection{From Max WSR to Min WSMSE}

For a general Rx filter $\bmf_k$ we have the MSE
\beq
\begin{array}{l}
e_k(\bmf_k,\bmg)
 = (1-\bmf_k^H\bmH_{k,b_k}\bmg_k)(1-\bmg_k^H\bmH_{k,b_k}^H\bmf_k)\\[1mm]
+\sum_{i\neq k}\bmf_k^H\bmH_{k,b_i}\bmg_i\bmg_i^H\bmH_{k,b_i}^H\bmf_k +   ||\bmf_k||^2 =1\!-\!\bmf_k^H\bmH_{k,b_k}\bmg_k\\[1mm]
 \!-\bmg_k^H\bmH_{k,b_k}^H\bmf_k\!+\!\!\dsum_{i}\!\bmf_k^H\bmH_{k,b_i}\bmg_i\bmg_i^H\bmH_{k,b_i}^H\bmf_k\! +\!   ||\bmf_k||^2 .
\vspace{-3mm}
\end{array}
\label{eqWSMSE0}
\eeq
The $WSR(\bmg)$ is a non-convex and complicated function of $\bmg$. 
Inspired by \cite{b5}, we introduced \cite{b6}, \cite{b7} an augmented cost function,
the Weighted Sum MSE, $WSMSE(\bmg,\bmf,w)$
\beq
 = \sum_{k=1}^K u_k(w_k\; e_k(\bmf_k,\bmg)-\ln w_k) + \sum_{i=1}^C\lambda_i (\sum_{k:b_k=i} ||\bmg_k||^2 \!-\! P_i)
\label{eqWSMSE1}
\eeq
where $\lambda_i$ = Lagrange multipliers.
After optimizing over the aggregate auxiliary Rx filters $\bmf$ and weights $w$, we get the WSR back:
\beq
\min_{\bmf,w} WSMSE(\bmg,\bmf,w) = - WSR(\bmg) +\sum_{k=1}^K u_k
\label{eqWSMSE2}
\eeq
The advantage of the augmented cost function: alternating optimization leads to solving simple quadratic or convex functions:
\beq
\mbox{}\!\!\!
\begin{array}{l}
\dmin_{w_k} WSMSE\;\Rightarrow\; w_k = 1/e_k\\[1mm]
\dmin_{\bmf_k} WSMSE\Rightarrow\bmf_k\! =\! (\sum_{i}\bmH_{k,b_i}\bmg_i\bmg_i^H\bmH_{k,b_i}^H \!+\!   I_{\Nr_k})^{-1}\bmH_{k,b_k}\bmg_k\\[1mm]
\dmin_{\bmg_k} WSMSE\,\Rightarrow \\[1mm]
\bmg_k\!=\! (\sum_{i}u_iw_i\bmH_{i,b_k}^H\bmf_i\bmf_i^H\bmH_{i,b_k} \!+\! \lambda_{b_k} I_{\Nt})^{-1}\bmH_{k,b_k}^H\bmf_ku_kw_k
\vspace{-2mm}
\end{array}
\label{eqWSMSE3}
\eeq
{\it UL/DL duality}: the optimal Tx filter $g_k$ is of the form of a MMSE linear Rx
for the dual UL in which $\lambda$ plays the role of Rx noise variance
and $u_kw_k$ plays the role of stream variance.

\subsection{Minorization (DC Programming)}

In a classical difference of convex functions (DC programming) approach (also called Successive Convex Approximation (SCA)) as in \cite{b8}, the concave signal terms  are kept and the convex interference terms are replaced 
by the linear (and hence concave) tangent approximation. This linearization is in term of the Tx covariance matrix $\bmQ_k$. 
However, after substituting $\bmQ_k=\bmG_k\bmG_k^H$ in terms of BF matrices $\bmG_k$, the concave character is less clear. But in any case, 
this DC programming/SCA approximation allows to construct a minorizer cost function, and minorization is a well established optimization approach \cite{b9}.

So, consider the dependence of WSR on $\bmQ_k$ alone.
Then
\beq
\begin{array}{l}
WSR = u_k \ln\det(\bmR_{\kbar}^{-1}\bmR_k) + WSR_{\kbar}\, ,\\[2mm]
WSR_{\kbar} = \sum_{i=1,\neq k}^K u_i \ln\det(\bmR_{\ibar}^{-1}\bmR_i) 
\end{array}
\label{eqKG2}
\eeq
where $\ln\det(\bmR_{\kbar}^{-1}\bmR_k)$ is concave in $\bmQ_k$ and $WSR_{\kbar}$ is convex in $\bmQ_k$.
Since a linear function is simultaneously convex and concave, consider the first order Taylor series expansion in $\bmQ_k$
around the current\footnote{To keep notation light, we shall not denote $\bmR_i$, $\bmA_{k}$ as $\bmR_i^{'}$, $\bmA_{k}^{'}$ etc.}
  $\bmQ^{'}$ (i.e.\  all $\bmQ^{'}_i$) with e.g.\   $\bmR_i=\bmR_i(\bmQ^{'})$, then
\beq\!\!\!\!
\begin{array}{l}
WSR_{\kbar}(\bmQ_k,\bmQ^{'}) \approx WSR_{\kbar}(\bmQ_k^{'},\bmQ^{'}) - \tr\{ (\bmQ_k-\bmQ^{'}_k) \bmA_k\}\\[2mm]
\bmA_{k} = \!-\left.\dfrac{\partial WSR_{\kbar}(\bmQ_k,\!\bmQ^{'}\!)}{\partial \bmQ_k}\right|_{\bmQ^{'}_k,\bmQ^{'}} \!\!\!\!\!= \!\!\!
\dsum_{i\neq k}^K \!\!\! u_i\bmH_{i,b_k}^H(\bmR_{\ibar}^{-1}\!\!\! -\! \!\bmR_{i}^{-1})\bmH_{i,b_k}
\end{array}
\label{eqKG3}
\eeq
Note that the linearized (tangent) expression for $WSR_{\kbar}$ constitutes a lower bound for it.
Now, dropping constant terms, reparameterizing the $Q_k = \bmG_k\bmG_k^H$, performing this linearization for all users, 
and augmenting the WSR cost function with the Tx power constraints,
we get the Lagrangian
\beq
\begin{array}{l}
WSR(\bmG,\bmG^{'},\bmlambda) =\dsum_{j=1}^{C}\lambda_j P_j+\\
 \dsum_{k=1}^K u_k \ln\det(1+\bmG_k^H \bmB_k\bmG_k)
-\bmG_k^H(\bmA_{k}+\lambda_{b_k} I)\bmG_k
\end{array}\!
\label{eqKG4}
\eeq
where
\beq
\bmB_k = \bmH_{k,b_k}^H\bmR_{\kbar}^{-1}\bmH_{k,b_k}\; .
\label{eqKG4b}
\eeq
The gradient (w.r.t.\  $\bmG_k$) of this concave WSR lower bound is actually still the same as that of the original WSR criterion!
And it allows an interpretation as a generalized eigenvector condition
\beq
 \bmB_k\,\bmG_k= (\bmA_{k}+\lambda_{b_k}I)\bmG_k\; \frac{1}{u_k} (I+\bmG_k^H\bmB_k\bmG_k)
\label{eqKG5}
\eeq
or hence 
$
\bmGb_k = V_{max}(\bmB_k, \bmA_{k}+\lambda_{b_k}I)
$ 
are the (normalized) "max" generalized eigenvectors of the two indicated matrices,
with eigenvalues $\Sigma_k = \Sigma_{max}(\bmB_k, \bmA_{k}+\lambda_{b_k}I)$.
Let $\Sigma_k^{(1)} = \bmGb_k^{H}\bmB_k\bmGb_k$ and
$\Sigma_k^{(2)} = \bmGb_k^{H}\bmA_k\bmG_k$. 
The advantage of formulation (\ref{eqKG4}) is that it allows straightforward power adaptation:
introducing stream powers in the diagonal matrices $\bmP_k\geq 0$ and 
substituting $\bmG_k = \bmGb_k\, \bmP_k^{\frac{1}{2}}$ in (\ref{eqKG4}) yields
\beq
\begin{array}{l}
WSR(\bmP,\bmlambda) =\sum_j^C \lambda_j P_j +\\
\dsum_{k=1}^K [u_k \ln\det(I+\bmP_k\Sigma_k^{(1)})-\tr\{\bmP_k(\Sigma_k^{(2)}+\lambda_{b_k} I)\}]
\end{array}
\label{eqKG6}
\eeq
optimization of which leads to the following interference leakage aware water filling (WF)
(jointly for the $P_k$ and $\lambda_c$)
\beq
\bmP_k = \lb u_k (\Sigma_k^{(2)}+\lambda_{b_k} I)^{-1} - \Sigma_k^{-(1)}\rb^+ , \;
\sum_{k: b_k=c}\!\!\! \tr\{\bmP_k\} = P_c
\label{eqKG7}
\eeq
where the Lagrange multipliers are adjusted to satisfy the power constraints. 
This can be done by bisection and gets executed per BS.
Note that some Lagrange multipliers could be zero.
Note also that as with any alternating optimization procedure, there are many updating schedules possible, with different impact on convergence speed.
The quantities to be updated are the $\bmgb_k$, the $\bmP_k$ and the $\lambda_c$.
Note that the minorization approach, which avoids introducing Rxs, can at every BF update allow to introduce an arbitrary number of streams per user
by determining multiple dominant generalized eigenvectors, and then let the WF operation decide how many streams can actually be sustained.

In contrast, in \cite{b8}, for given $\bmlambda$, the $\bmG$ get iterated till convergence and
the $\bmlambda$ are found by duality (line search):
\beq
\begin{array}{l}
\displaystyle
\min_{\bmlambda\geq 0} \max_{\bmG} [\sum_j^C \lambda_j P_j +\sum_k \{u_k \ln\det(\bmR_{\kbar}^{-1}\bmR_k)-\lambda_{b_k} \tr\{\bmP_k\} \} ] \\
= 
\displaystyle
\min_{\bmlambda\geq 0} WSR(\bmlambda) .\vspace{-2mm}
\end{array}
\label{eqKG8}
\eeq
This typically leads to higher computational complexity for a given convergence precision.

\subsection{Pathwise Wireless MIMO Channel Model}

In this section we drop the user index $k$ for simplicity.
The MIMO channel transfer matrix at any particular subcarrier $n$ of a given OFDM symbol can be written as 
\cite{b10},\cite{b11}
\vspace{-2mm}
\beq
\begin{array}{l}
\bmH[n] = \sum_{i=1}^L A_i e^{j\psi_i[n]}\bmh_r(\phi_i)\bmh_t^T(\theta_i) = \bmH_r\; \bmPsi[n]\;\bmD\;\bmH_t^H \; ,\\[5mm]
\bmH_r = \lsb\bmh_r(\phi_1)\,\bmh_r(\phi_1)\cdots\rsb\, ,\,\\[3mm]
\bmPsi[n] = \!\!\lsb\!\!\begin{array}{ccc} \! e^{j\psi_1[n]}\!\!\!\! & & \\ & \!\!\!\!e^{j\psi_2[n]}\!\!\!\! & \\ & & \!\!\ddots\!\end{array}\!\!\rsb\!\!
 , 
\bmD = \!\!\lsb\!\!\begin{array}{ccc} \!A_1\!\!\!\! & & \\ & \!\!\!\! A_2\!\!\! & \\ & & \!\!\!\ddots \!\end{array}\!\!\rsb\!\! ,
\bmH_t^H = \!\!\lsb\!\!\!\begin{array}{c} \bmh_t^T(\theta_1)\\ \bmh_t^T(\theta_2)\\ \vdots\end{array}\!\!\!\rsb
\end{array}
\label{eqCSIT1}
\eeq
where there are $\Np$ (specular) pathwise contributions with
\begin{itemize}\itemsep 0mm
\item
$A_i>0$: path amplitude
\item
$\psi_i[n]$: path phase
\item
$\theta_i$: angle of departure (AoD)
\item
$\phi_i$: angle of arrival (AoA)
\item
$\bmh_t(.)$/$\bmh_r(.)$: $\Nt / \Nr \times 1 $ Tx/Rx antenna array response
\end{itemize}
with $||\bmh_t(.)||=1$, $||\bmh_r(.)||=\Nr$.
For wideband scenarios, all factors may become frequency-dependent.
The antenna array responses are just functions of angles AoD, AoA in the case of standard antenna arrays with scatterers in the far field. 
The fast variation of the phases $\psi_i$ (due to Doppler)  corresponds to the fast fading.
All the other parameters vary on a slower time scale and correspond to slow fading.
In the pathwise CSIT (pwCSIT) model, we shall assume the  $\psi_i$ to be i.i.d. uniformly distributed and all
slow parameters to be known.
Note that the pathwise channel model, which leads here to a type of Tx covariance CSIT, does not lead to the usual separable covariance case, which is discussed e.g. in \cite{b2}.
In previous work, we essentially modeled the whole of $\bmH_r \bmPsi$ as i.i.d. random, which leads to a special case of the MIMO channel with separable correlation structure. Here the knowledge of $\bmH_r$ is exploited, leading to an appearance of (implicit) Rxs
who contribute to the interference management.

\begin{figure}[htb]\centering
\includegraphics[width=1\columnwidth]{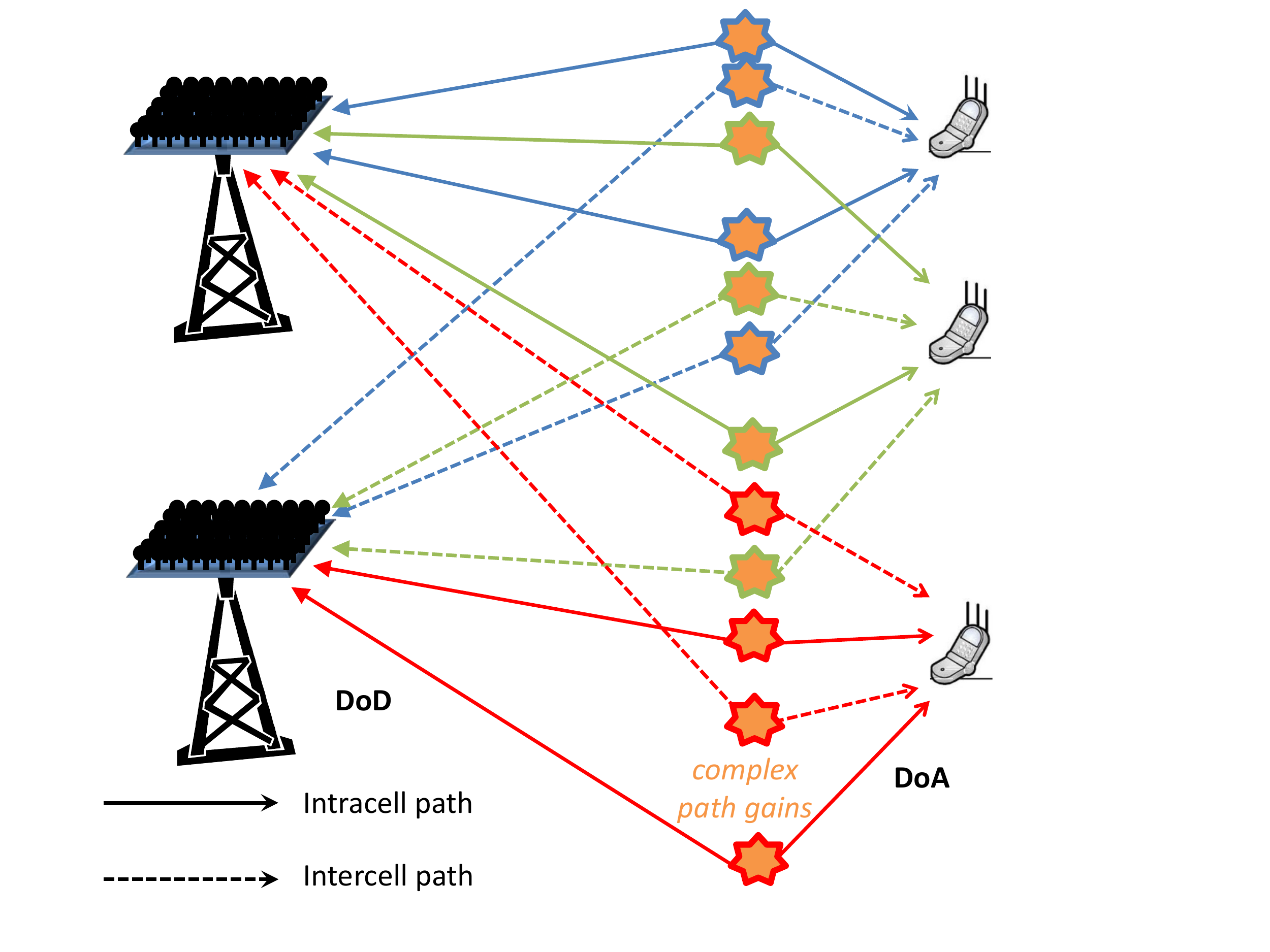}\vspace{-5mm}
\caption{Pathwise Multi-User Multi-Cell scenario.}
\label{FigPathwiseIBC}
\vspace{-5mm}
\end{figure}

\section{MIMO Interference Alignment (IA)}

ZF (IA) feasibility for both the general reduced rank MIMO channels case and the pathwise MIMO case has been discussed in \cite{b3}, in particular also when only based on Tx side covariance CSIT. It is shown how the IA responsability gets shared between Tx and Rx, requiring only local CSI.
Also the role of Rx antennas is highlighted, leading to reduced (Tx covariance) rank channels.

\section{Expected WSR (EWSR)}

For the WSR criterion, we have assumed so far that the channel $\bmH$ is known. The scenario of interest however is that of partial CSIT.
Once the CSIT is imperfect, various optimization criteria could be considered, such as outage capacity. Here we shall consider the expected
weighted sum rate $\E_{\bmH|\bmHb} WSR(\bmg,\bmH) =$
\beq
EWSR(\bmg) =  \E_{\bmH|\bmHb} \sum_k u_k\ln(1 + \bmg_k^H\bmH_{k,b_k}^H \bmR_{\kbar}^{-1} \bmH_{k,b_k}\bmg_k)
\label{eqEWSR1}
\eeq
where we now underlign the dependence of various quantities on $\bmH$ and $\bmHb$ is a channel estimate.
The EWSR in (\ref{eqWSR1}) corresponds to perfect CSIR since the optimal Rx filters $\bmf_k$ as a function of the aggregate $\bmH$ have
been substituted, namely $WSR(\bmg,\bmH) = \max_{\bmf}\sum_k u_k(-\ln (e_k(\bmf_k,\bmg)))$.

In the MaMIMO limit, we obtain the {\em Massive EWSR limit} in which
\beq
\E_{\bmH|\bmHb} \ln\det(I + \bmH\bmQ\bmH^H) \rightarrow \ln\det(I + \E_{\bmH|\bmHb}\{\bmH\bmQ\bmH^H\})
\label{eqEWSR2}
\eeq
when $\Nt\rightarrow\infty$ for finite $\Nr$.
The gap between both sides in \eqref{eqEWSR2} can be analyzed and is bounded for any MIMO size by $\gamma$ (Euler-Mascheroni) in the worst case of only a single Rayleigh fading entry. The RHS also corresponds to the Expected Weighted Sum Unbiased MSE (EWSUMSE) approach introduced in \cite{b12}, which is a useful formulation by itself. The RHS also becomes the exact mutual information if we consider Gaussian channel outputs instead of Gaussian channel inputs.

For the case of mean (channel estimate) and covariance CSIT being jointly captured by the Gaussian CSIT, 
$\mbox{vec}(\bmH^T) = \bmh \sim \CN(\bmhb,C_{\bmh\bmh})$ where $\bmhb = \mbox{vec}(\bmHb^T)$, we get
\beq
\E\{ \bmH \bmg\bmg^H \bmH^H\} = \bmHb \bmg\bmg^H \bmHb^H + (I_{\Nr} \otimes \bmg^T) C_{\bmh\bmh} (I_{\Nr} \otimes \bmg^*)\, .
\label{eqEWSR3}
\eeq
This general paCSIT model, even with a pathwise channel model, could account for unmodeled paths, estimation errors on the path parameters, etc. Here we shall consider that all paths are modeled and perfectly known, except for the path phases.

\subsection{Massive EWSR with pwCSIT}

For the special case of pwCSIT \eqref{eqCSIT1} considered here, if the total number of paths (all users) becomes very large, the path phases average out 
and by the law of large numbers
\beq
\begin{array}{l}
\E_{\Psi} \ln\det(I + \bmH \bmQ\bmH^H) \approx \ln\det(I + \E_{\Psi} \bmH \bmQ\bmH^H)\\
\bmH\bmQ\bmH^H \longrightarrow
\E_{\Psi}\, \bmH\bmQ\bmH^H 
 = \bmH_r \, \bmD\, diag(\bmH_t^H \bmQ \bmH_t)\, \bmD\, \bmH_r^H
\end{array}
\label{eqEWSR4}
\eeq
which is now frequency-independent, and where $diag(.)$ denotes the diagonal matrix obtained by taking the diagonal part of the matrix argument.
Hence we get the following MaMIMO limit matrices
\beq\!\!
\begin{array}{l}
\bmR_k[n]\! = \!\bmI_{\Nr_k}\! + \!\!\dsum_{i=1}^K \bmH_{r,k,b_i}  \bmD_{k,b_i}^2\, diag(\bmH_{t,k,b_i}^H \bmQ_i \bmH_{t,k,b_i}) \bmH_{r,k,b_i}^H
 \\[2mm]
\bmR_{\kbar}[n]\! =\! \bmI_{\Nr_k}\! +\!\! \dsum_{i\neq k} \bmH_{r,k,b_i} \bmD_{k,b_i}^2\, diag(\bmH_{t,k,b_i}^H \bmQ_i \bmH_{t,k,b_i}) \bmH_{r,k,b_i}^H
\end{array}
\label{eqEWSR5}
\eeq
This leads to e.g. (with $\bmQ_i = \bmG_i\bmG_i^H$) :
$
\frac{\partial \ln\det(\bmR_k)}{\partial \bmG_i^*} \! = \!
\bmH_{t,k,b_i} diag(\bmH_{r,k,b_i}^H \bmR_k^{-1} \bmH_{r,k,b_i})\bmD_{k,b_i}^2 \bmH_{t,k,b_i}^H \bmG_i$
and we can introduce
\[\!\!
\begin{array}{l}
\bar{\bmB}_k =
\bmH_{t,k,b_k}\, diag(\bmH_{r,k,b_k}^H \bmR_{\kbar}^{-1} \bmH_{r,k,b_k})\, \bmD_{k,b_k}^2\, \bmH_{t,k,b_k}^H\\[2mm] 
\bar{\bmA}_k \!= \!\!\dsum_{i\neq k}^K\!\! u_i
\bmH_{t,i,b_k} diag(\bmH_{r,i,b_k}^H (\bmR_{\ibar}^{-1}\!\!\!-\!\bmR_i^{-1}) \bmH_{r,i,b_k}) \bmD_{i,b_k}^2 \bmH_{t,i,b_k}^H 
\end{array}
\]
It suffices now to replace the matrices $\bmA_k$, $\bmB_k$ in the minorization approach by the matrices
$\bar{\bmA}_k$, $\bar{\bmB}_k$ above to get a maximum EWSR design:
$
\bmG_k^{'} = V_{max}(\bar{\bmB}_k, \bar{\bmA}_k+\lambda_{b_k} I).$
With $\Sigma_k^{(1)} = \bmG_k^{'H}\bar{\bmB}_k\bmG_k^{'}$,
$\Sigma_k^{(2)} = \bmG_k^{'H}\bar{\bmA}_k\bmG_k^{'}$, $\bmG_k = \bmG_k^{'}\, \bmP_k^{\frac{1}{2}}$,
\beq
\bmP_k = \lb u_k (\Sigma_k^{(2)}+\lambda_{b_k} I)^{-1} - \Sigma_k^{-(1)}\rb^+  , \;
\sum_{k:b_k=c} \tr\{\bmP_k\} = P_c\; .
\vspace{-5mm}
\label{eqEWSR9}
\eeq

\subsection{Interference management by Tx/Rx}

Interference management by Tx:\vspace{-2mm}
\beq\!\!\!
\bmR_k[n]\! = \!\bmI_{\Nr_k}\! + \!\!\dsum_{i=1}^K\! \bmH_{r,k,b_i}  \bmD_{k,b_i}^2 diag(\underbrace{\bmH_{t,k,b_i}^H \bmG_i}_{}
\underbrace{\bmG_i^H \bmH_{t,k,b_i}}_{}) \bmH_{r,k,b_i}^H
\label{eqEWSR10}
\eeq
where the underbraced terms would be zero for $i\neq k$ in case of a ZF design (high SNR optimal).
One can identify {\bf implicit Rxs}: from $\bar{\bmA}_k $ we get
\beq\!\!
\begin{array}{l}
diag(\bmH_{r,i,b_k}^H (\bmR_{\ibar}^{-1}\!-\!\bmR_i^{-1}) \bmH_{r,i,b_k})\\[2mm]
= diag(\bmH_{r,i,b_k}^H\! \underbrace{\bmR_i^{-1} \bmH_{r,i,b_i}}_{\bmF_i}(\!\bmD_i \!- \!\bmH_{r,i,b_i}^H \!\bmR_i^{-1}
\bmH_{r,i,b_i}\!)^{-1}\\[2mm]
\bmH_{r,i,b_i}^H\!\bmR_i^{-1}\bmH_{r,i,b_k})
= diag(\underbrace{\bmH_{r,i,b_k}^H \bmF_i }_{}\widetilde{\bmD}_i\underbrace{\bmF_i^H\bmH_{r,i,b_k}}_{})
\end{array}
\vspace{-2mm}
\label{eqEWSR11}
\eeq
where the $\bmF_i$ are implicit Rxs and again the underbraced terms would be zero for $i\neq k$ in case of a ZF design.

\subsection{iCSIT vs pwCSIT WSR at low SNR}

We have $WSR=$
\beq
\dsum_{k=1}^K\! u_k \ln\det(I+\bmG_k^H\bmH_{k,b_k}^H\bmF_k(\bmF_k^H\bmR_{\kbar}\bmF_k)^{-1}\bmF_k^H\bmH_{k,b_k}\bmG_k) 
\label{eqEWSR12}
\eeq
where
$\bmR_{\kbar} = I$ for low SNR (or high SNR below). 
At low SNR, the optimal Tx/Rx are matched filters. We get :\\
WSR at low SNR for iCSIT
\beq
WSR=
 \!\dsum_{k=1}^K\! u_k \ln\det(I + \Sigma^2(\bmH_{k,b_k})P_k)
\label{eqEWSR13}
\eeq
and 
WSR at low SNR for pwCSIT : $WSR = $
\beq
\dsum_{k=1}^K\! u_k \ln\det(I + \bmH_{r,k,b_k}^H\!\bmH_{r,k,b_k}  \bmD_{k,b_k}^2 diag(\bmH_{t,k,b_k}^H \bmQ_k \bmH_{t,k,b_k}))
\vspace{-3mm}
\label{eqEWSR14}
\eeq

\subsection{iCSIT vs pwCSIT WSR at high SNR}

With again the WSR in \eqref{eqEWSR12},\\
WSR at high SNR for iCSIT : \eqref{eqEWSR12} where the $\bmG$, $\bmF$ satisfy $\;\bmF_k^H\bmH_{k,b_i}\bmG_i = 0\, , \; i\neq k$
which reflects joint Tx/Rx ZF.
.\\
On the other hand, 
WSR at high SNR for pwCSIT :\vspace{-2mm}
\[
\bmH_{r,k,b_i} = [\underbrace{\bmH_{r,k,b_i,r}}_{\mbox{by UE}} \; \underbrace{\bmH_{r,k,b_i,t}}_{\mbox{by BS}} ]  , \;
\bmH_{t,k,b_i} = [\underbrace{\bmH_{t,k,b_i,r}}_{\mbox{by UE}} \; \underbrace{\bmH_{t,k,b_i,t}}_{\mbox{by BS}} ]
\]
where the underbraces indicate which nodes handle the interference of the indicated channel portions, and 
\vspace{-1mm}
\[
\begin{array}{l}
\bmF_k = P^{\perp}_{\bmH_{r,k,r}}\bmH_{r,k,b_k}\, (\bmH_{r,k,b_k}^HP^{\perp}_{\bmH_{r,k,r}}\bmH_{r,k,b_k})^{-\frac{1}{2}}\\
\bmG_k^{'} = P^{\perp}_{\bmH_{t,k,t}}\bmH_{t,k,b_k}\, (\bmH_{t,k,b_k}^HP^{\perp}_{\bmH_{t,k,t}}\bmH_{t,k,b_k})^{-\frac{1}{2}}\\
\!\!\!
WSR=
 \!\dsum_{k=1}^K\! u_k \ln\det(I+\Sigma(\bmS_k^{\frac{1}{2}}\, \bmD_{k,b_k}^2\; diag\{\bmT_k\}\bmS_k^{\frac{1}{2}})\,P_k)\\
\bmS_k = \bmH_{t,k,b_k}^H P^{\perp}_{\bmH_{t,k,t}}\bmH_{t,k,b_k}\, , \;
\bmT_k = \bmH_{r,k,b_k}^H P^{\perp}_{\bmH_{r,k,r}}\bmH_{r,k,b_k} .
\end{array}
\]
In the pathwise case, the ZF task of all paths gets split between Tx and Rx which each ZF paths from either Tx or Rx side.

\section{Simulation Results\vspace{-2mm}}

Simulations are provided for the case of $C=2$ cells,  2 users/cell,  $\Np=3$ paths in all channels,
and varying Tx/Rx antenna numbers $\Nt$, $\Nr$. The expected sum rate is compared between the cases of perfect instantaneous CSIT (iCSIT)
and (global) pathwise CSIT (pwCSIT). The loss is limited as soon as pathwise ZF is possible.
\vspace{-4mm}

\begin{figure}[htb]
  \centerline{\includegraphics[width=0.9\columnwidth]{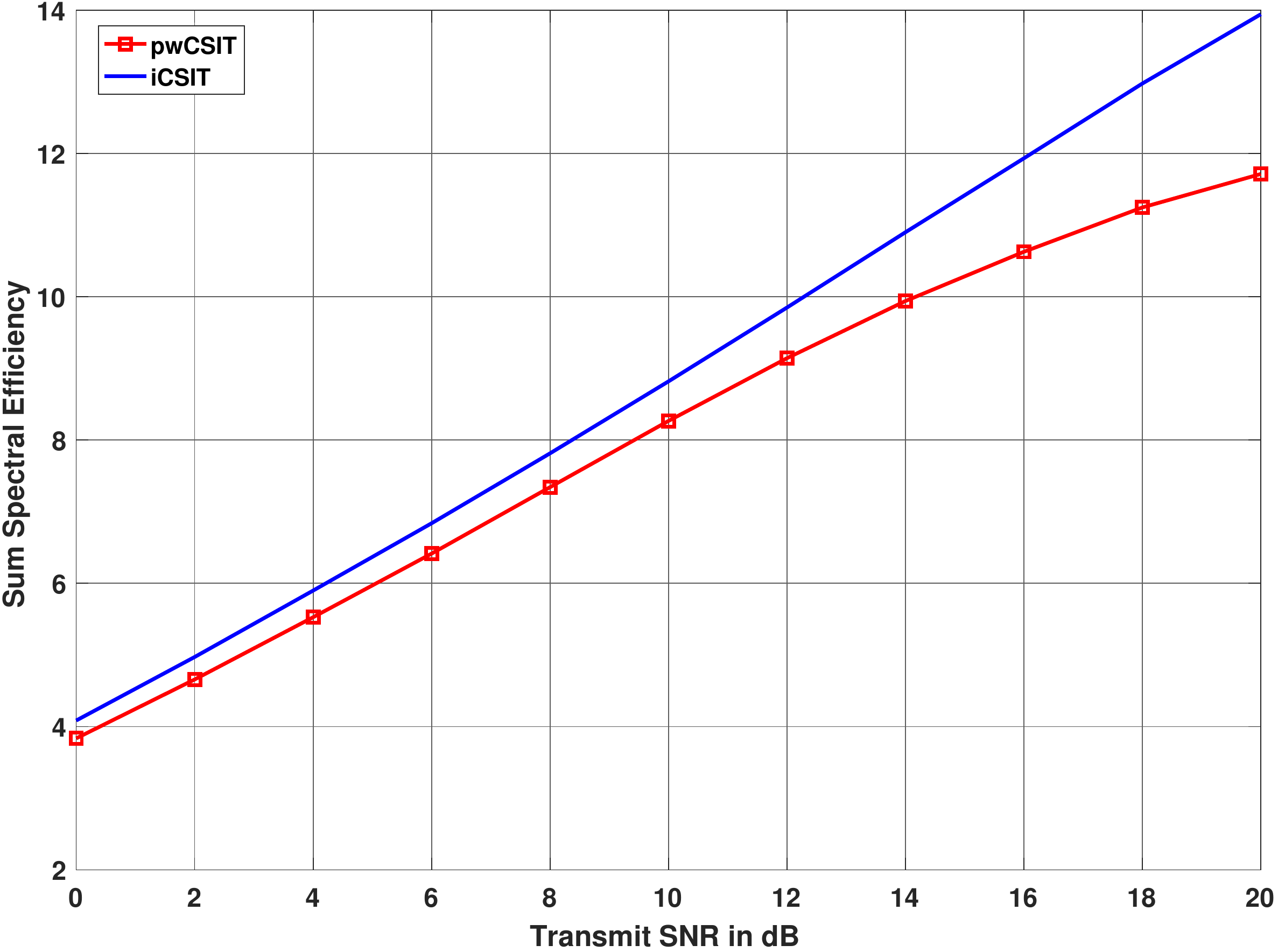}}
 \vspace{-4mm}
\caption{Expected sum rate comparison for $\Nt=3, \Nr=3 $.}
\vspace{-6mm}
\end{figure}

\begin{figure}[htb]
  \centerline{\includegraphics[width=0.9\columnwidth]{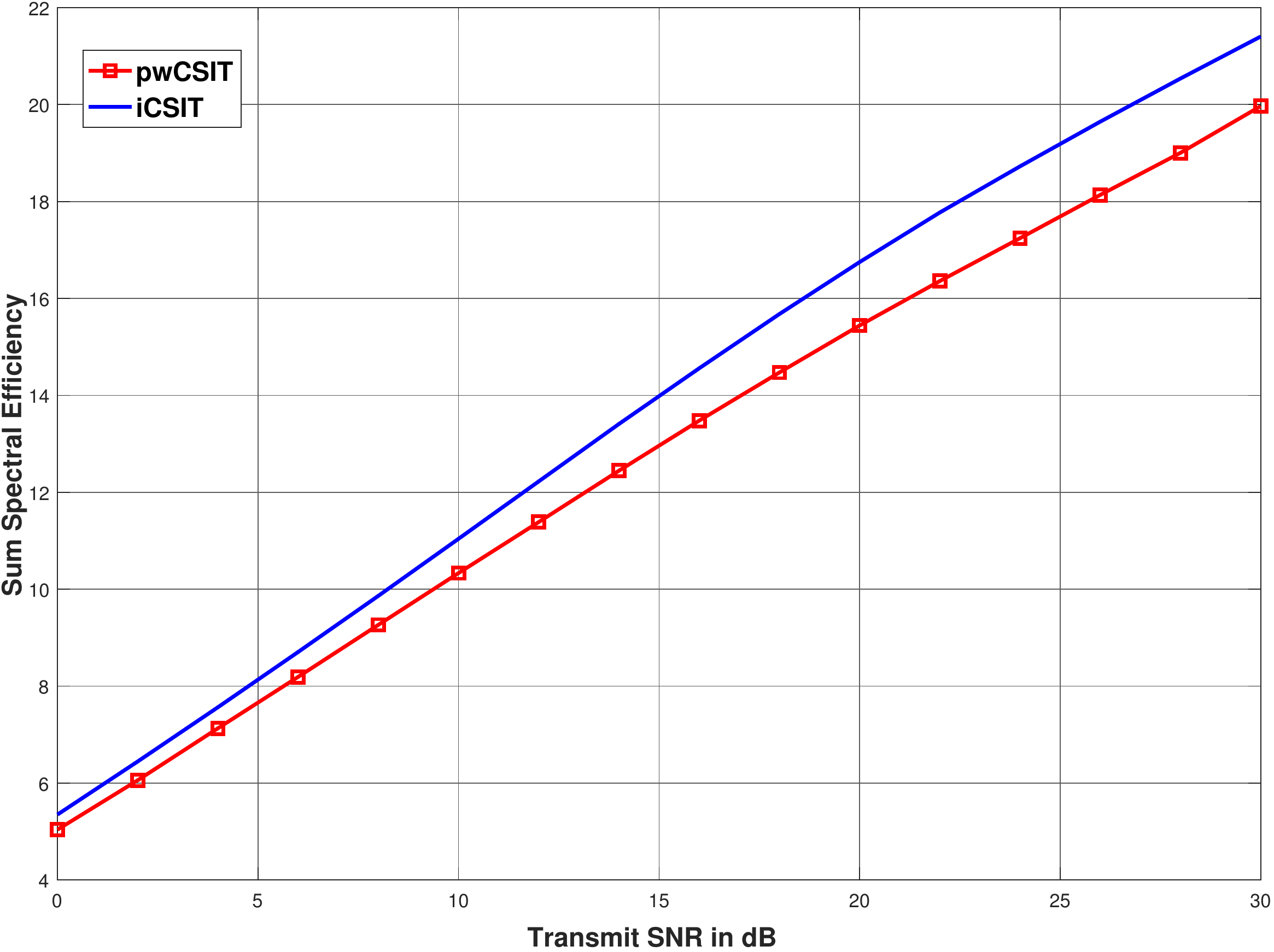}}
 \vspace{-4mm}
\caption{Expected sum rate comparison for $\Nt=4, \Nr=4 $.}
\vspace{-8mm}
\end{figure}

\begin{figure}[htb]
  \centerline{\includegraphics[width=0.9\columnwidth]{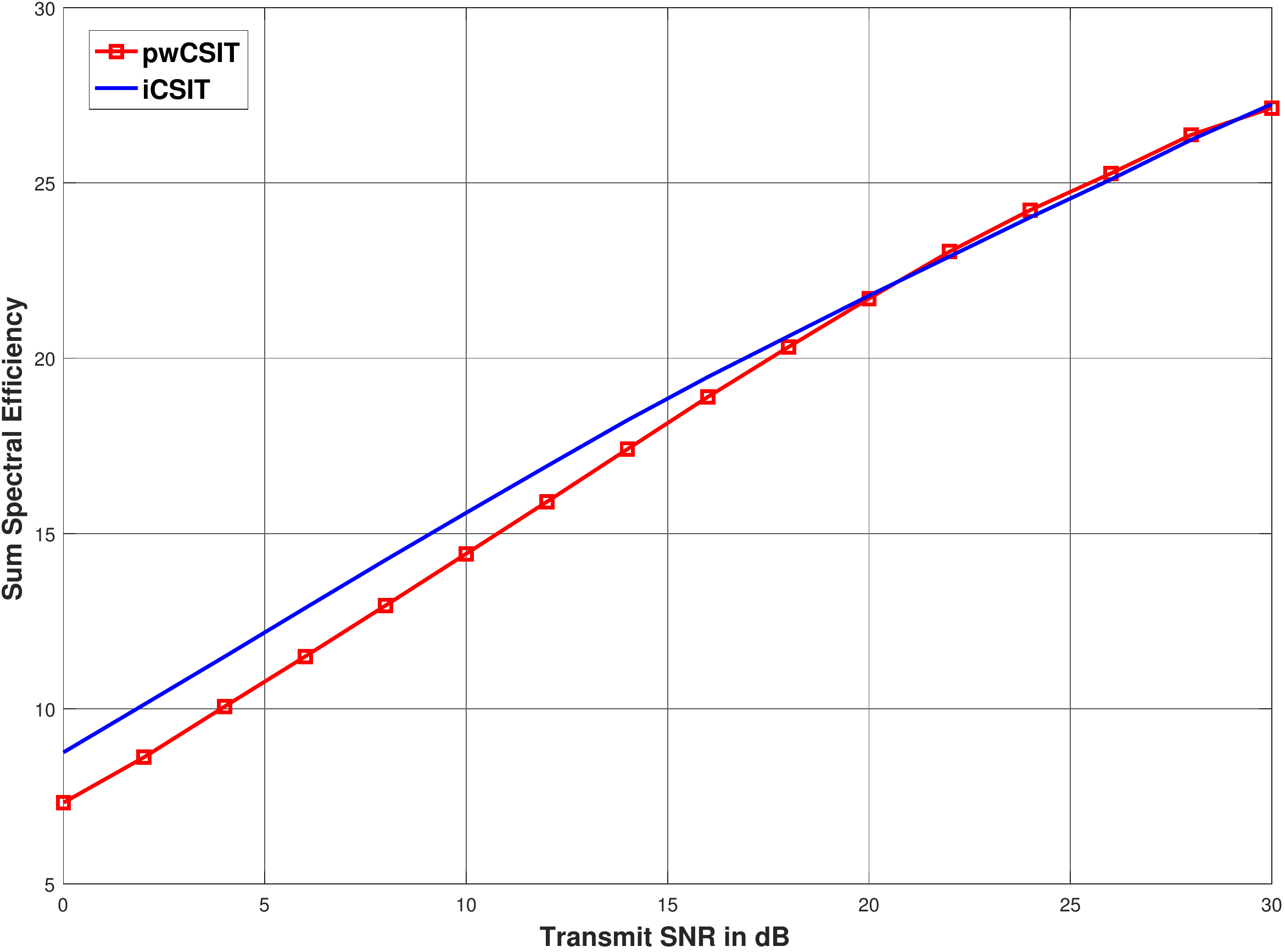}}
 \vspace{-3mm}
\caption{Expected sum rate comparison for $\Nt=10, \Nr=4 $.}
\vspace{-6mm}
\end{figure}

\section*{Acknowledgments}

EURECOM's research is partially supported by its industrial members:
ORANGE, BMW, SFR, ST Microelectronics,
Sy\-man\-tec, SAP, Monaco Telecom, iABG,  and by the projects HIGHTS (EU H2020)
and MASS-START (French FUI).
The research of Orange Labs is partially supported by the EU H2020 project One5G.


\end{document}